\documentclass[aps,prl,preprint,showpacs,preprintnumbers,amsfonts,a4paper,byrevtex]{revtex4}

\newcommand{\be}{\begin{eqnarray}}
\newcommand{\ee}{\end{eqnarray}}
\newcommand{\ben}{\begin{eqnarray*}}
\newcommand{\een}{\end{eqnarray*}}

\usepackage{graphics}
\usepackage{epsf}
\usepackage{epsfig}

\begin{document} 

%\flushbottom
\setcounter{page}{1}

\title{Topological Origin of the Phase Transition in a Model of DNA Denaturation}

\author{Paolo Grinza}
\email{grinza@sissa.it}

\author{Alessandro Mossa}
\email{mossa@sissa.it}

\affiliation{Istituto Nazionale di Fisica Nucleare (INFN) and International School for Advanced Studies (SISSA),\\  via Beirut 2-4, I-34014 Trieste, Italy}

%\date{29\,September, 2003}
%\date{Received 12\,December, 2002, published 12\,June 2003}

\begin{abstract}
The hypothesis that phase transitions originate from some topological change of the critical level hypersurface of the potential energy receives direct evident support by our study of the Bishop--Peyrard model of DNA thermal denaturation.
\end{abstract}

% 05.70.Fh = Phase transitions: general studies
% 02.40.Pc = General Topology
% 63.70.+h = Statistical mechanics of lattice vibrations and displacive phase transitions
% 87.10.+e = General theory and mathematical aspects (biological/medical physics)


\preprint{SISSA 85/2003/FM}

\thispagestyle{empty}

\maketitle

%\section{Introduction}

In spite of the huge body of literature on the subject of thermodynamic phase transitions, one has at least two reasons to not be completely satisfied with our present understanding of such prominent phenomenon: From a theoretical viewpoint, the loss of analyticity in the derivatives of the free energy \cite{Yang:be} and the consequent appearance of singularities in the thermodynamic observables are only macroscopic signals of the phase transition, while the physical origin should be looked for at the microscopical dynamics level \cite{bib:G03}. On the other hand, transitional phenomena experimented in finite atomic clusters, spin glasses, and biological systems still challenge physicists to find out new tools and methods to cope with their difficulties.          

Both aspects of the problem are addressed by a recently proposed geometric approach to phase transitions \cite{Caiani:1997yv,Franzosi:1999cq,bib:CPC00}. Given any Hamiltonian system, the configuration space can be endowed with a metric, in order to obtain a Riemannian geometrization of the dynamics \cite{bib:CPC00}. At the beginning, several numerical and analytical studies of a variety of models showed that the fluctuation of the curvature becomes singular at the transition point. Then in Ref.~\cite{Caiani:1997yv} the following  conjecture was proposed: The phase transition is determined by a change in the topology of the configuration space, and the loss of analyticity in the thermodynamic observables is nothing but a consequence of such topological change. The latter conjecture is also known as the topological hypothesis. 

The aim of this Letter is to shed some other light upon the actual mechanism underlying the topological change. In order to do this, we choose to study a system which, though one dimensional (1D), has the peculiarity to exhibit a second order phase transition, the Peyrard-Bishop model \cite{bib:PB89}, developed as a simple model for describing the DNA thermally induced denaturation. Its simplicity already permitted an analytical computation of the largest Lyapunov exponent \cite{bib:BD01} by exploiting the geometrical method proposed in Ref.~\cite{Caiani:1997yv}. On the other hand, we pursued a study of the topological characterization of the phase transition in a full analytic way. The Letter is organized as follows: First, we give a brief description of the main features of the Peyrard-Bishop model, the main body is devoted to the study of the topology of the configuration space of the model, and, last, we discuss our results and state our conclusions.

\section{The model}

The system we studied was introduced in 1989 by Peyrard and Bishop \cite{bib:PB89} as a simple model for DNA thermal denaturation. It is defined by the following Hamiltonian
\be \label{model}
H  =  \sum_{i=1}^N \bigg[ {p_i^2\over2m}+{K\over 2}(y_{i+1}-y_i)^2   
 +   D(e^{-ay_i}-1)^2+   Dhay_i \bigg] \,,
\ee
which represents the energy of a string of $N$ base pairs of reduced mass $m$. Each hydrogen bond is characterized by the stretching $y_i$ and its conjugate momentum $p_i=m(dy_i/dt)$. The elastic transverse force between neighboring pairs is tuned by the constant $K$, while the energy $D$ and the inverse length $a$ determine, respectively, the plateau and the narrowness of the on-site potential well that mimics the interaction between bases in each pair. It is understood that $K$, $D$, and $a$ are all positive parameters. The transverse, external stress $h\ge0$ is a computational tool useful in the evaluation of the susceptibility. Our interest in it lies in the fact that a phase transition can occur only when $h=0$. We assume periodic boundary conditions.

%\begin{figure} 
%\begin{center}
%\includegraphics[width=7cm,height=4.95cm]{fig1_b.eps}
%\caption{The Morse potential.}
%\end{center}
%\end{figure} 
The transfer operator technique
\cite{bib:SSF72,bib:KS75}
maps the problem of computing the classical partition function into the easier task of evaluating the lowest energy eigenvalues of a ``quantum'' mechanical Morse oscillator (no real quantum mechanics is involved, since the temperature plays the role of $\hbar$). One can then observe that, as the temperature increases, the number of levels belonging to the discrete spectrum decreases, until for some critical temperature $T_c=2\sqrt{2KD}/(ak_B)$ only the continuous spectrum survives. This passage from a localized ground state to an unnormalizable one corresponds to the second order phase transition of the statistical model. Various critical exponents can be analytically computed and all applicable scaling laws can be checked. Detailed calculations as well as references to original works can be found in Ref.~\cite{bib:DTP02}. 

\section{Topology of configuration space}

According to the Morse theory \cite{bib:M}, it is possible to understand the topology of a given manifold by studying the regular critical points of some smooth function (called Morse function) defined on it. In our case, the manifold $\mathcal{M}$ is the configuration space $\mathbb{R}^N$ and the natural choice for the Morse function is the potential $V(y)$. Hence, one is lead to define the following family of submanifolds of $\mathcal{M}$:
\be
{\mathcal M}_v=\{y \in {\mathbb R}^N:V(y) \leq v \} \,.
\ee
A full characterization of the topological properties of ${\mathcal M}_v$ generally requires one to find the critical points of $V(y)$, which means solving the equations
\be \label{critcond}
   {\partial V\over \partial y_i}=0, \qquad \qquad i=1,\dots,N \,.
\ee
Moreover, one has to compute the indexes of all the critical points, that is the number of negative eigenvalues of the Hessian $\partial^2V/(\partial y_i\partial y_j)$.
Then the Euler characteristic $\chi({\mathcal M}_v)$ can be computed by means of the formula
\be \label{chiemu}
   \chi({\mathcal M}_v)=\sum_{k=0}^N (-1)^k\mu_k({\mathcal M}_v) \,,
\ee
where $\mu_k({\mathcal M}_v)$ is the total number of critical points of $V(y)$ on ${\mathcal M}_v$ which have index $k$. If applied to any generic model, such a task turns out to be quite formidable, but the exceptional simplicity of the Peyrard--Bishop model (\ref{model}) makes it possible to carry on completely the topological analysis without invoking Eq.~(\ref{chiemu}).

For the potential in exam, Eq.~(\ref{critcond}) results in the nonlinear system
\be \label{2ordfindiff}
   {a\over R}(y_{i+1}-2y_i+y_{i-1})=h-2(e^{-2ay_i}-
   e^{-ay_i}) \,,
\ee
where $R=Da^2/K$ is a dimensionless ratio. It is easy to verify that a particular solution is given by
\be \label{minimum}
   y_i=-{1\over a}\ln{1+\sqrt{1+2h}\over 2} \qquad \qquad i=1,\dots,N \,.
\ee
The corresponding minimum of potential energy is
\be
   V_{\mathrm{min}}=ND\left({1+h-\sqrt{1+2h}\over2}-h\ln{1+\sqrt{1+2h}\over
   2}\right) \,.
\nonumber
\ee
Actually, there is only one other solution. Here is the proof: Equation (\ref{2ordfindiff}) can also be interpreted as a finite difference second order equation for the evolution of the variable $y_i$. Let us introduce a set of new variables $x_i=(y_i-y_{i-1})$. We can fix an initial condition $y_N=y_0=\bar{y}$ so that this definition can be reversed into $y_i=\bar{y}+\sum_{j=1}^ix_j$. In terms of these new variables, Eq.~(\ref{2ordfindiff}) is written
\be \label{recursive}
   x_{i+1}=x_i  +  {R\over a}\Bigg[h-2\exp \Bigg(-2a\bar{y}-2a\sum_{j=1}^i
   x_j\Bigg)
 +2\exp\Bigg(-a\bar{y}-a\sum_{j=1}^ix_j\Bigg)\Bigg]
\ee
while the periodicity condition reads simply $\sum_{j=1}^N x_j=0$. Now we prove
that no solution is compatible with this requirement besides the trivial solution
$x_i=0 \ \forall i=1,\dots,N$. From the recursive relation Eq.~(\ref{recursive}),
we deduce that $x_{i+1}\ge x_{i}$ if and only if
\be
   \exp\left(-a\bar{y}-a\sum_{j=1}^{i} x_j\right)\le{1+\sqrt{1+2h}\over 2} \,.
\ee
Let us suppose that for some $k$ it results
$x_k>0$. If $x_k \ge x_{k-1}$, then \emph{a fortiori} $x_{k+1} \ge x_k$ because $e^{-ax_k}<1$.
Now $x_{k+1}>0$ and $x_{k+1}\ge x_k$ imply $x_{k+2} \ge x_{k+1}$ and so on:
The series $\{x_j\}$ increases for $j>k$.
On the other hand, if $x_k<x_{k-1}$, then $e^{-ax_{k-1}}<1$; hence, $x_{k-1}<x_{k-2}$
and so on. Either situation is incompatible with a periodic boundary condition.
 A completely analogous
argument rules out the possibility that it could be $x_k<0$ for some value
of $k$.

We have proven that no solution is possible for Eq.~(\ref{2ordfindiff})
unless it results $y_i=Y$ $\forall i$. Now it is evident that, when $h=0$, the limiting point $Y\longrightarrow +\infty$ is a critical point. Here it is useful
to introduce new variables $w_i\equiv e^{-ay_i}$, so that the condition of vanishing
gradient reads
\be
   {1\over R}\ln{w_i^2\over w_{i+1}w_{i-1}}=h-2w_i(w_i-1) \,,
\ee
from which we immediately recognize, beside the minimum Eq.~(\ref{minimum}), a
second solution $w_i=0 \ \forall i=1,\dots,N$, present only when $h=0$. This
is an irregular critical point because its Hessian matrix is degenerate; hence,
the potential can no longer be considered a good Morse function. Two remarks are in order: First, as already stated, we do not need Eq.~(\ref{chiemu}) because ${ \mathcal M}_v$ is homotopous to a full disk; second, since the Euler
characteristic is a homotopical invariant,
it does not change at $v_c$ for the submanifolds ${\mathcal M}_v$.  

Notwithstanding this, we still have a way to complete our analysis. Let us consider the \emph{boundaries} of $M_v$, the hypersurfaces
\be
\Sigma_v \equiv \{y \in {\mathbb R}^N : V(y)=v \} \,.
\ee
If we choose one direction $\tilde{y}=(\tilde{y}_1,\dots,\tilde{y}_N)$ and study the
asymptotic behavior of the function $V(y)$ in that direction by taking the limit
$\lambda\longrightarrow\infty$ in $V(\lambda\tilde{y})$, we find that if there is
at least one negative component $\tilde{y}_j<0$, then the limit is positive infinity because
of the exponential term. If all of the components are positive, then the potential goes as
\be
   {K\lambda^2\over2}\sum(\tilde{y}_{i+1}-\tilde{y}_i)+\lambda
   D h a\sum\tilde{y}_i+ND \,,
\ee
so that the limit is always positive infinity unless all $\tilde{y}_j$ are
equal and $h=0$. This implies that for $v$ raising from $V_{\mathrm{min}}$ to
$v_c=ND$ it happens that $\Sigma_v$ is a
closed ($N$--$1$) hypersurface [$\chi({ \Sigma}_v)=0$ or 2 depending on $N$ being
odd or even, respectively].
\begin{figure}\label{hypersurf}
\begin{center}
\epsfig{figure=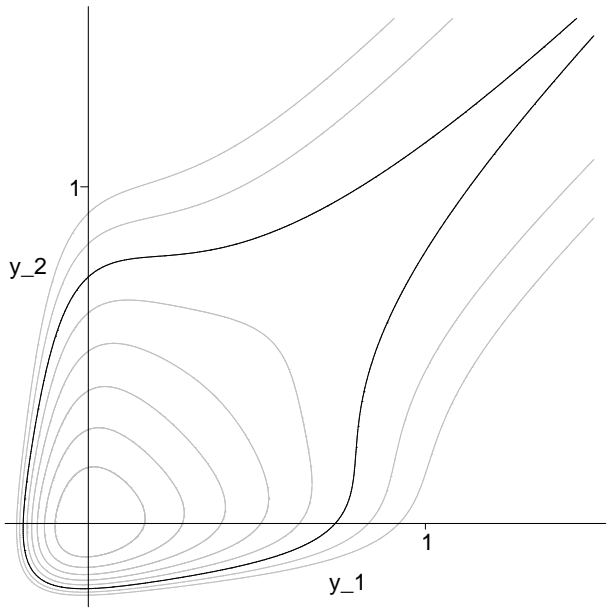,width=8cm}
\caption{Level hypersurfaces, $N=2$. The black line is the critical hypersurface.}
\end{center}
\end{figure}
On the contrary, if $v=v_c$ and $h=0$, in the direction
singled out by the condition $\tilde{y}_1=\tilde{y}_2=\dots=\tilde{y}_N$,
the level hypersurface
$\Sigma_{v_c}$ closes at positive infinity (see Fig.~\ref{hypersurf} for a $N=2$ example), while for each $v>v_c$ the level hypersurfaces fail to close [$\chi({ \Sigma}_v)=1$]. This topological change
is evidently related to the phase transition,
since its presence is due to the very same feature of the potential (its plateau)
giving rise to the phase transition.

\section{Conclusions}

Since its original formulation \cite{Caiani:1997yv}, the topological hypothesis has received several indirect and direct confirmations. If we limit ourselves to the latter, we may cite the numerical study of the $\phi^4$ theory \cite{Franzosi:1999cq} and the analytical computations of the Euler characteristic of the mean-field XY model \cite{bib:ccp99,bib:CPC03} and of the mean-field $k$-trigonometric model \cite{bib:ACPRZ03}.

The main difference between our work and previous results lies in the low dimensionality of the Peyrard-Bishop model. By choosing a 1D system, we get an enormous simplification of the landscape of critical points of the potential $V(y)$, resulting in a completely transparent description of the relation between topology and phase transition. Indeed, the same mechanism responsible for the appearance of the phase transition is clearly seen to produce the topology change in configuration space. 

Moreover, the present work contributes to enlarge the record of cases (quite a useful job in a research field so recent and full of open questions such as this) in at least two respects: It is the first analytical study of the topology of a \emph{noncompact} configuration manifold and, while previous results \cite{bib:ACPRZ03} identify the phase transition point in a discontinuity of the first derivative of $\chi({\mathcal M}_v)$, the present work is peculiar in the sense that the topological change is signaled by the Euler characteristic of the boundary submanifolds $\chi(\Sigma_v)$.
 
In conclusion, it should be remarked that, though the Peyrard--Bishop model is very simple, it represents a first step in the exploration of the complex and interesting world of biological transition phenomena. Topological techniques will hopefully provide valuable insight to such a challenging research topic.
 
\begin{acknowledgments}
This work was partially supported by the European Commission TMR programme HPRN-CT-2002-00325 (EUCLID). The work of P. G. is supported by the COFIN ``Teoria dei Campi, Meccanica Statistica e Sistemi Elettronici''.
\end{acknowledgments}

\end{document}